\documentclass[onecolumn]{emulateapj}

\shorttitle{Galaxies rotation}
\shortauthors{C. E. Navia}

\usepackage{amssymb}	

\begin{document}


\title{
On the rotation curves of galaxies at low and high redshifts
}


\author{C. E. Navia\altaffilmark }
\affil{Instituto de Fisica, Universidade Federal Fluminense, 24210-346, Niteroi, Rio de Janeiro, Brazil}


\altaffiltext{1}{E-mail address:navia@if.uff.br}


\begin{abstract}
The thermal history of the Universe is included in the Debye Gravitational Theory (DGT) to describe the speed of rotation of the galaxies. The DGT incorporate the temperature of Debye in the entropic gravitational theory. According to the DGT, the expression of the rotation of the galaxies is not a continuous function of the redshift, there is a discontinuity at $\sim 0.77$. According to their redshift, the galaxies form two groups. (a) Those with redshift above 0.77 with declining rotation curves like $R^{\beta}$ with $\beta \leq -0.5$ (Newtonian regime). (b) Those with redshift below 0.77 with rising rotation curves like $R^{\beta}$ with $\beta \geq 0$ (Mondian regime). At $z\sim 0.77 $ an extra boost (Dirac delta-like twisting force) led the galaxies to spin very fast.
 This scenario is consistent with a mysterious entity 
such as the dark energy and that at $z\sim 0.77 $, in addition to accelerating the expansion, boosted the rotation of the galaxies again, inducing a transition between the Newtonian regime to the Mondian regime. It is possible to check that the characteristics on galaxy rotations provided by the DGT for a broad range of redshift from $z\sim 4$ to $z=0$ are in agreement with the observations. We believe that the change in the form of rotation of galaxies at $z\sim 0.77$ is new evidence for the hypothesis of dark energy.
\end{abstract}






\section{Introduction}
\label{sec:intro}

Condensed matter physics is one the of the central disciplines of
20th-century physics. Remarkable success is reached with the construction of effective models describing solids regarding
fictitious new entities (quasiparticles and collective excitations).
These new methods make it possible not only to solve a large number of
problems, and also to obtain many new relations of a general character.
This remarkable success make of the solid state physics become source of models to other fields of physics, 
such as the quantum field theory.

In 1932 Tamm introduced the concept the `` phonon'' \citep{zim01} a quasi-particle as a quantum unit of sound.
The main application of phonons considered as quantum vibrational states is in the description of thermal properties, the speed of sound through the materials, very useful in the study of the specific heat of solids at low temperatures. Indeed, the ``phonon'' concept was inspired by the Debye \citep{deb12} theory of the specific heat of solid at low temperatures. The Debye model correctly predicts the low-temperature dependence of the heat capacity of solid and coincides with the approaching the Dulong-Petit law at high temperatures. 

It is possible to introduce the structure of Debye in the thermodynamic theories of gravity, as in Verlinde emergent gravity theory \citep{ver11}, this allows for the formulation of Debye's Gravitational Theory (DGT). The number of bits of information stored on the holographic screen, play the role of  ``phonons'' and their vibrational states follows a continuous range de frequencies that depend on temperature. This picture is useful and allows incorporating the thermal history of the Universe, for example, in the relation of the rotation speed of the galaxies. 

The DGT has two asymptotic boundaries, starting with a Newtonian regime at high temperatures (high redshifts) and ending with a Mondian regime at low temperatures (low redshifts). A mysterious unknown entity induces the transition between these two boundaries and everything leads one to believe that it is related to that obtained by observing supernovae of type 1A and that supports the hypothesis of dark energy.

We would like to point out, that the dark energy hypothesis is not assumed a priori but based on the results inferred from the analysis of the rotation velocity of the galaxies, that is, the effects of the dark energy are embedded in the theory.

The connection between thermodynamic and gravity began in the 70s \citep{bek73} and \citep{haw74}, researching the nature of black holes. \citep{jac95} using the  Davies-Unruh effect \citep{dav75,unr76} shown a thermodynamic description of gravity obtaining the Einstein's equations.
Also, \citep{pad15} suggests that the association between gravity and entropy leads in a natural way to describes gravity as an emergent phenomenon, and a formalism of gravity as an entropic force is derived by \citep{ver11}. The dependence of information on the surface area, rather than volume (Holographic principle) \citep{hoo93}, it is one of the keys of black hole thermodynamic theory, as well as in DGT.

This paper is organized as follow: Section~\ref{sec:gravity} is devoted to the background of the DGT theory.
Section~\ref{sec:rotation} presents a straightforward analysis of galaxy rotation curves.
This section includes a description of the thermal history of the Universe, based on the cosmic microwave background (CMB) temperature variation with the redshift.
The cosmological implications of our results are described in section 
\ref{sec:cosmo}. This section includes a straightforward analysis by the DGT to determine the expansion acceleration as a function of redshift. There is a  temporal coincidence between the beginning of the accelerated expansion and the beginning of the sudden jump in the angular velocity driven by an external twisting force.
Finally, section~\ref{sec:conclusion} is devoted to conclusions and some comments.

 \section{The basis of the Debye Gravitational theory (DGT)}
 \label{sec:gravity}

In 1912, \citep{deb12} developed a theory to explain the heat capacity of 
solid as low temperatures. He assumed that the vibration of the atoms of 
the lattice of a solid follows a continuous rIn 1912, \citep{deb12} 
developed a theory to explain the heat capacity of solid 
as low temperatures. He assumed that the vibration of the atoms of the 
lattice of a solid follows a continuous range of frequencies, such as an 
elastic structure, that cuts off at a maximum frequency, $\omega_D$. 
In this theory, each solid has a specific temperature, called as Debye 
temperature, linked with the cutting frequency by the relation 
$T_D=\hbar \omega D/k_B$, where $\hbar$ and  $k_B$ are the reduced Planck 
constant and the Boltzmann constant, respectively.
The Debye theory correctly predicts the dependence of the heat capacity 
of solids with the temperature in a wide range of temperatures, 
including the very low temperatures and coincides with the approaching t
he Dulong-Petit law at high temperatures.

\begin{figure}
\vspace*{-2.0cm}
\hspace*{0.0cm}
\centering
\includegraphics[width=16.0cm]{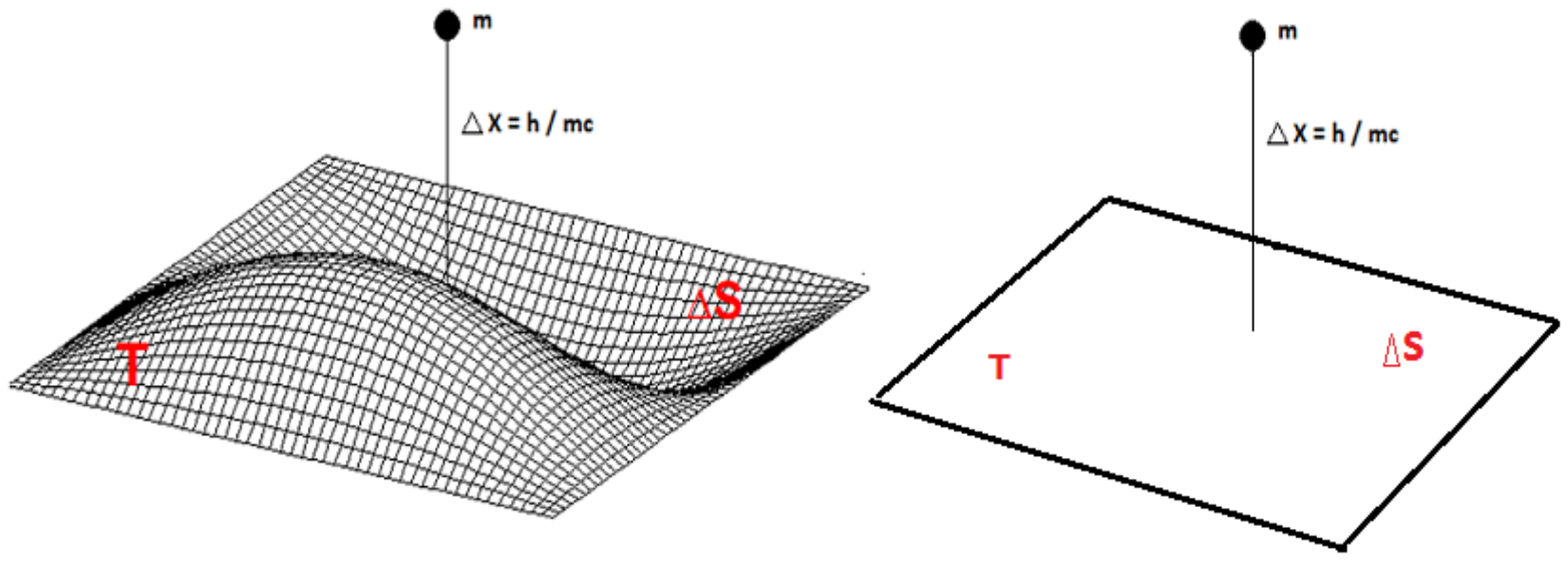}
\vspace*{-12.5cm}
\caption{Scheme of a particle of mass m close to a holographic screen   
that stores information about a system. This approximation produces a variation of the entropy on the screen and causes a force to emerge on the particle. The left panel consider a holographic screen like an elastic oscillating membrane (DGT picture), whereas the right panel, the holographic screen is considered only as a smooth surface, without oscillations (Verlinde picture) and plays the role of Dulong-Petit law of the theory of solids and it is observed only at high temperatures.
}
 \label{fig:tela}
\end{figure}

In DGT a holographic screen is the closed area where is stored the information 
of the matter within the screen. The information is codified in N bits and is 
considered as the freedom degrees of the system. The holographic screens 
coincide with the Newtonian equipotential surfaces. 

A central argument of the holographic principle considers that each bit of 
information on the screen carries an energy given by $1/2k_BT$ and according 
to the holographic principle, the number of bits N on the screen surface is 
proportional to the area of the screen. These N bits constitute the number o
f freedom degrees of the system. Taking into account the equipartition of 
energy principle, the specific potential energy on the screen can be written a
\begin{equation}
U=\frac{1}{2}N k_B T. 
\label{eq:energy1}
\end{equation}
Following the analogy with Debye model, the expression of the specific potential energy to low temperatures can be obtained substituting $k_BT$, for $k_BT\;\mathcal{D}_1(T_D/T)$.
\begin{equation}
U=\frac{1}{2}N k_B T \mathcal{D}_1\left(\frac{T_D}{T}\right).
\label{eq:energy2}
\end{equation}
The main difference with the Debye theory of solids is that the third 
Debye function, $\mathcal{D}_3(x)$ was replaced by the first Debye 
function, $\mathcal{D}_1(x)$. In the Debye theory of solids, the ``phonons'' 
have three vibrational states, two transverses and one longitudinal, already 
in the case of the bits on the holographic screen
 they have only the longitudinal vibrational state because the holographic 
screen is an equipotential surface.
As in the Debye model, it is assumed that the vibrations follow a continuous 
range of frequencies, which cuts off at a maximum frequency $\omega_D$. The 
definition of $\mathcal{D}_1$ is 
\begin{equation}
\mathcal{D}_1\left(\frac{T_D}{T}\right)= \left(\frac{T}{T_D}\right) \int_0^{T_D/T} \frac{x}{e^x-1}dx,
\label{eq:dbye1}
\end{equation}
The shape of Debye function reflects the Bose-Einstein statistic formula, used in its derivation. 
 If M represents all mass enclosed by the holographic screen surface, the specific potential energy can be written as $U=Mc^2$. Replacing this expression in equation~\ref{eq:energy2}, we have a relation to the temperature as
\begin{equation}
T \mathcal{D}_1\left(\frac{T_D}{T}\right)=\frac{2Mc^2}{Nk_B}.
\label{eq:temperature}
\end{equation}  
  
Considering that the entropy variation of the information, $\Delta S$ on the holographic screen, happens when a particle of mass m is at a distance $\Delta X$
(close to the Compton wavelength). The Bekestain entropy variation can be expressed as

\begin{equation}
\Delta S=2\pi k_B \frac{mc}{\hbar} \Delta X.
\label{eq:entropy}
\end{equation} 
Fig.~\ref{fig:tela} shows a schematic representation of equation~(\ref{eq:entropy}). Its is the Bekenstein-Verlinde representation of a particle of massa m close to a holographic screen, in the case of right panel, the information ``bits'' stored on the screen, do not have vibrational states. Already in the left panel is shown the same scheme, however, in this case  the bits on the screen have a vibrational state and follows  a continuous range of frequencies, such as an elastic membrane, and it is the  case that is explored in this paper.
Equation~(\ref{eq:entropy}), allows to obtain the entropic force defined as 
\begin{equation}
F=T \frac{\Delta S}{\Delta x}.
\label{eq:force}
\end{equation}
Considering that the number (N ) of bits on the  holographic screen  is given by
$N=(c^3/G\hbar)A$, where A is the area of the screen and taking into account that
$a=F/m$ gives the acceleration of the mass m.  The more simple case is for a system with spherical symmetry, and considering the holographic screen as a sphere of radius R, ($A=4\pi \;R^2$). Combining equations~\ref{eq:force}, 
equation~\ref{eq:entropy} and  equation~\ref{eq:temperature}  one gets that 

\begin{equation}
a\mathcal{D}_1\left(\frac{T_D}{T}\right)= \frac{GM}{R^2},
\label{eq:aceleration}
\end{equation}
that has two asymptotic limits


$$\mathcal{D}_1\left(\frac{T_D}{T}\right)=\left\{\begin{array}{rc}
1,&\mbox{if}\quad T_D/T < 1,\\
\frac{\pi^2}{6}(T/T_D), &\mbox{if}\quad T_D/T>1.
\end{array}\right.
$$

The first case (high T) reproduce the Newtonian dynamic, whereas the second limit (low T) can see used to obtain the only free parameter of the theory, the Debye temperature, $T_D$. As in the case of the study of the capacity specific of the solids at low temperatures, the obtention of Debye temperature can also be from observations. Also, an expression to the second asymptotic limit of the Debye function to $T_D/T>1$ is
\begin{equation}
\frac{\pi^2}{6}(\frac{T}{T_D})= \frac{a}{a_0},
\label{eq:debye2}
\end{equation}
where $a_0$ is a constant  acceleration scale linked to $T_D$. The equation~\ref{eq:debye2}
gives a bond between the temperature and acceleration and in this limit the equation~\ref{eq:aceleration}
become
\begin{equation}
a(\frac{a}{a_0})=\frac{GM}{R^2}.
\label{eq:mond}
\end{equation} 
For other temperature values outside this asymptotic limit, the equation~\ref{eq:mond} can be written in general as
\begin{equation}
a\left(\frac{a}{a_0}\right)^{\alpha}= \frac{GM}{R^2},
\label{eq:egt2}
\end{equation}
with the constraint condition
\begin{equation}
\mathcal{D}_1\left(\frac{a_0}{a}\right)=\left(\frac{a}{a_0}\right)^{\alpha}=
\left(\frac{\pi^2}{6}\frac{T}{T_D}\right)^{\alpha},
\label{eq:power}
\end{equation}
and which allows obtaining an expression for the index $\alpha$ as
\begin{equation}
\alpha=\frac{\log \mathcal{D}_1\left(\frac{a_0}{a}\right)}{\log \left(\frac{a}{a_0}\right)}=
\frac{\log \mathcal{D}_1\left(\frac{a_0}{a}\right)}{\log \left(\frac{\pi^2}{6} \frac{T}{T_D}\right)}.
\label{eq:alpha}
\end{equation}
 The equation~\ref{eq:egt2} has the same asymptotic limits than equation~\ref{eq:aceleration}
 

$$\left(\frac{a}{a_0}\right)^{\alpha}=\left\{\begin{array}{rc}
1,&\mbox{if}\quad \alpha =0 \;\;(T_D/T < 1),\\
\frac{a}{a_0}, &\mbox{if}\quad \alpha=1 \;\;(T_D/T>1).
\end{array}\right.
$$

The first asymptotic limit coincides with the Newtonian regime, while, the second asymptotic limit ($\alpha=1$)  coincides with the deep-MOND regime \citep{mil83,mil83b,san90},
and reproduce the empirical Tully-Fisher relationship observed between the mass M or intrinsic luminosity of nearby spiral galaxies and its asymptotic velocity of rotation. Considering the equation~\ref{eq:egt2} to $\alpha=1$ and taking into account that $a=\mathrm{v}^2/R$, the galaxy mass can be written as $M= (1/Ga_0)\mathrm{v}^4$ and plotted as $\log M=4 \mathrm{v} +\log (1/Ga_0)$ \citep{mcg11,fam12}.  The slope, 4,  fall precisely with that observed in galaxies at very low redshift, whereas the normalization requires $a_0=10^{-10}ms^{-2}$ \citep{mcg11}. The acceleration $a_0$ is the Milgrom acceleration parameter \citep{mil83}, and in this limit, DGT coincides with the deep-MOND regime. 

The asymptotic limit at low temperatures can provide also determining the only one free parameter of the DGT the Debye temperature $T_D$. It can be obtained as a function of the $a_0$ taking into account the bond between the acceleration and temperature expressed in equation~\ref{eq:debye2} and the $T_D$ is then
\begin{equation}
T_D=\frac{\pi^2 a_0}{6} (\frac{T}{a}),
\end{equation}
the ratio $(T/a)$ is provided by the Unruh effect
$T/a=\hbar/(2\pi K_B)$ and $T_D$ can be expressed as
\begin{equation}
T_D=\frac{\pi \hbar}{12k_Bc}a_0.
\label{eq:td}
\end{equation}

\begin{figure}
    \vspace{-0.0cm}
    \centering
	\includegraphics[width=12cm]{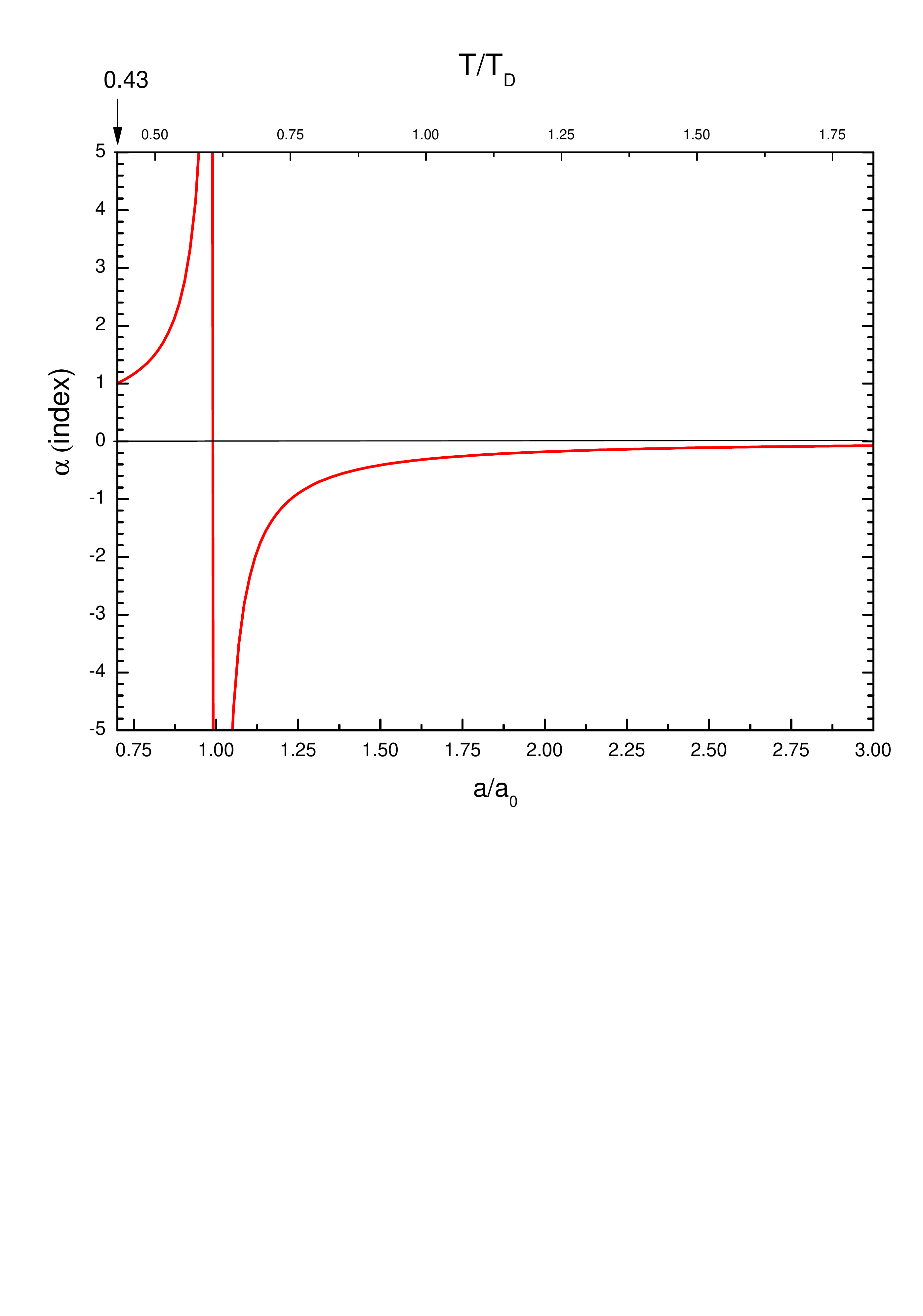}
	\vspace{-5.0cm}
    \caption{The $\alpha$ index 
as a function of the acceleration (bottom scale) and temperature (top scale).
The vertical arrow at top left corner, represent the temperature ratio  $T/T_D$ at $\alpha=1$. This value is useful in the calibration of the temperature of the holographic screen with the redshift.}
    \label{fig:alpha}
\end{figure}

Figure~\ref{fig:alpha} shows the dependence of the $\alpha$ index with acceleration (temperature), according to 
equation~\ref{eq:alpha}.
From this figure,  we can see that the transition between the two asymptotic limits,
from the deep-MOND regime ($\alpha=1$) to Newtonian regime ($\alpha=0$), has two regions well defined. In the area in which $a/a_0< 1$ the index $\alpha$ is positive and it is equal or higher than one. While in the area in which $a/a_0>1$ the index $\alpha$ is negative and asymptotically goes to zero. 

In addition there is a vertical asymptote at $a/a_0=1$, that is, the index $\alpha$ is a discontinuous function of
$a/a_0$ and diverge at $a/a_0=1$, the left-hand side limit of the $\alpha$ index goes to  
$\infty$ and the right-hand side limit of $\alpha$ index goes to
$-\infty$. 

\begin{figure}
    \vspace{-1.0cm}
    \centering
	\includegraphics[width=12cm]{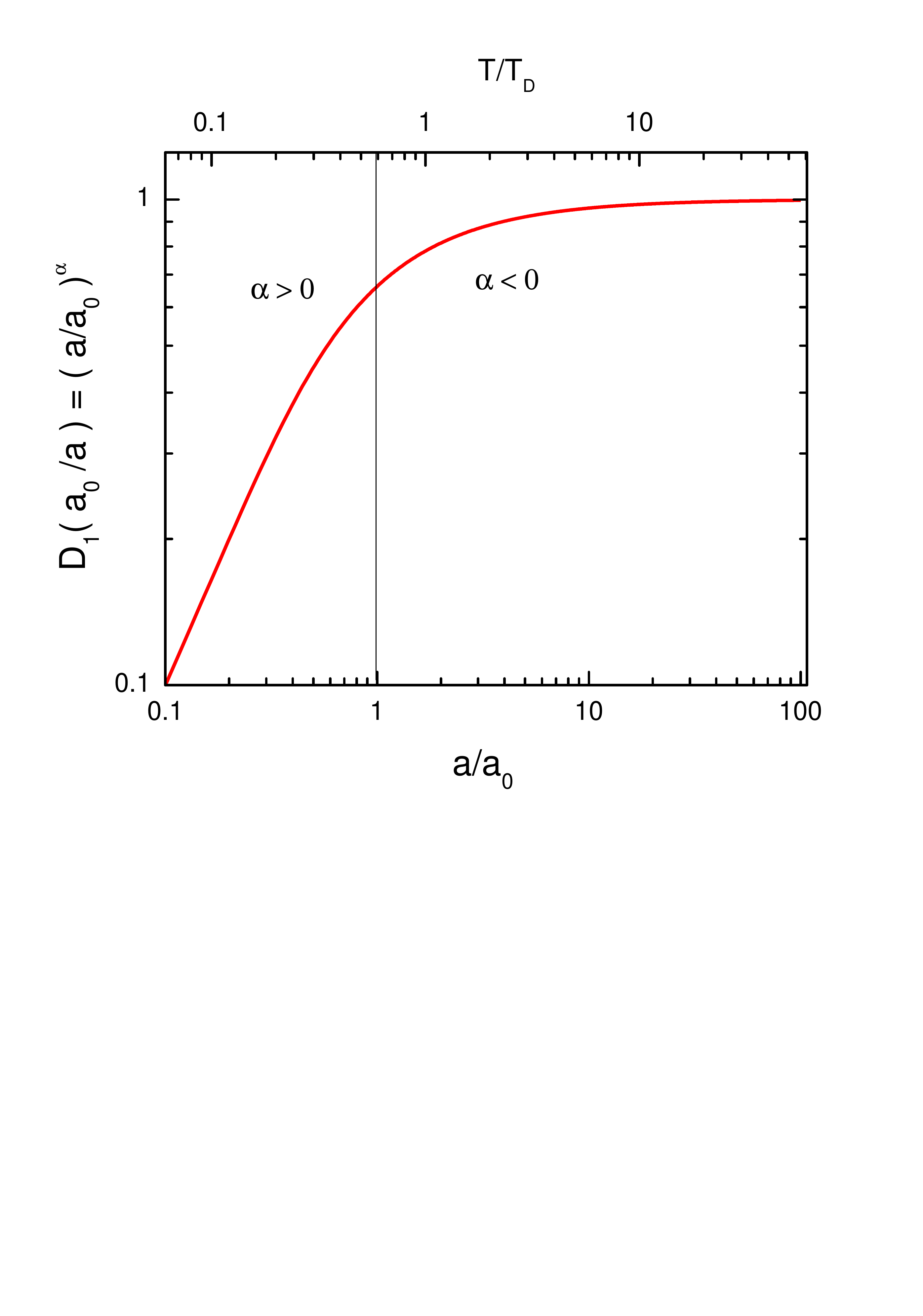}
	\vspace{-5.0cm}
    \caption{Debye first function, as a function of the acceleration (bottom scale) and temperature (top scale),  parametrized as a power law. Notice that in this panel, both abscissa and ordinate are in logarithmic scale.}
    \label{fig:power_law}
\end{figure}

However, this divergence has not influence in the Debye first function, because it is a continuous function of $a/a_0$. Figure~\ref{fig:power_law} shows a parametrization of the Debye first function as a power law as a function of the ratio $a/a_0$ (bottom scale) and $T/T_D$ (top scale).

 We would like to point out, that a power law function is an invariant scale relationship between two quantities, as well as, of natural physical interpretation. The equation~\ref{eq:egt2} is the key equation in the DGT, for systems with spherical symmetry.

\begin{figure}
    \vspace{-0.0cm}
    \centering
	\includegraphics[width=12cm]{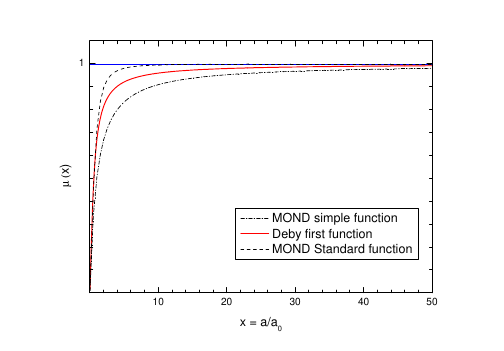}
	\vspace{-0.5cm}
    \caption{Comparison of the Debye first function and the two
    interpolating empirical functions more used in MOND, they are the simple and standard interpolating functions.
    }
    \label{fig:compara}
\end{figure}

 In DGT the Debye first function plays the role of interpolating function of the MOND theory, in this case, is no longer empirical. The main characteristic is that in DGT there in a bond between acceleration and temperature,  
this feature is extremely useful for our objectives. A figure~\ref{fig:compara} is a comparison of the Debye function and two expressions for the interpolating function, such as the simple and standard interpolating functions extensively used in MOND \citep{fam12}.

\section{Galaxy rotation curves: a straightforward analysis}
\label{sec:rotation}

\begin{figure}
    \vspace{-0.0cm}
    \centering
	\includegraphics[width=14cm]{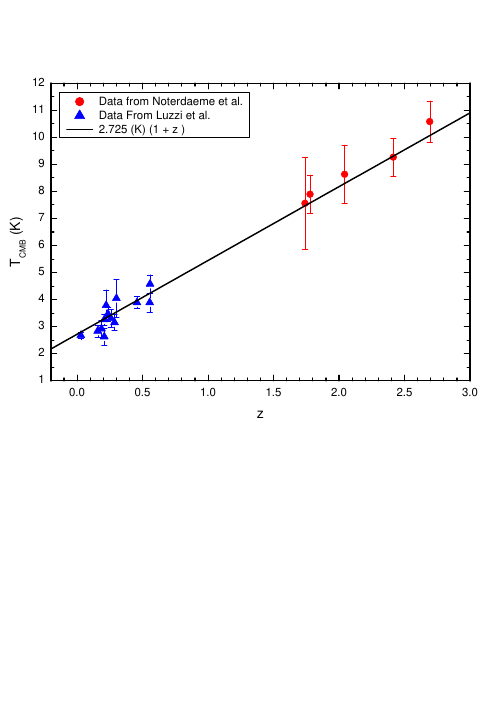}
	\vspace{-7.5cm}
    \caption{The black-body temperature of the Cosmic Microwave Background radiation as a function of redshift. Blue rectangles represent data from (Luzzi et al.( 2009)), and the red filled circles represent data from (Noterdaeme et al.( 2011)). The black solid line represents the evolution of $T_{CMB}$ as $(1+z)$.
}
    \label{fig:t_z}
\end{figure}

One of the profound observations of the twentieth century is that the universe is expanding, and implies that the universe was smaller, denser, and warmer in the distant past. A proof of this distant past is the cosmic microwave background  (CMB) radiation, and it is the residual radiation that's left when matter separated from the photons in the early stages of the Universe. Thus, the CMB cools as the Universe expands. As we follow the CMB on cosmological time, we must deal with the redshift z. Due to  CMB be similar to a black body radiation, the expression between the temperature of the CMB and redshift is $ T / T_0 = (1 + z) $, with $ T_0 = 2.725 $ K and it is the temperature of the CMB at present. This dependence of the CMB temperature with the red shift is by the astrophysical measurements \citep{luz09} (low redshift) and \citep{not11} (high redshift), as shown in the figure~\ref{fig:t_z}. That is evidence that indicates that, as the universe expands, it gets colder.

Let's beginner the analysis on the rotation curves of galaxies by the DGT,  indicating that it provides a bond between the acceleration and temperature, expressed in the equation~\ref{eq:power}. Consequently, there is also a bond between the acceleration and the redshift and the first Debye function could be parametrized as a function of the redshift. It is equivalent to obtain a relation between the index $\alpha$ and the redshift, as follows:

1.- We consider that for a given system, the temperature variation of the holographic screen with the redshift, follows the same dependence as observed in the CMB

\begin{equation}
\frac{T}{T_D} \propto (1+z).
\label{eq:ratio}
\end{equation}

2.- The holographic screen temperature at $z=0$ correspond  to $\alpha=1$ (deep-MOND regime)   and following the 
figure~\ref{fig:alpha} this occurs at 
$T/T_D \sim 0.43$. 

\begin{figure}
    \vspace{-0.0cm}
    \centering
	\includegraphics[width=12cm]{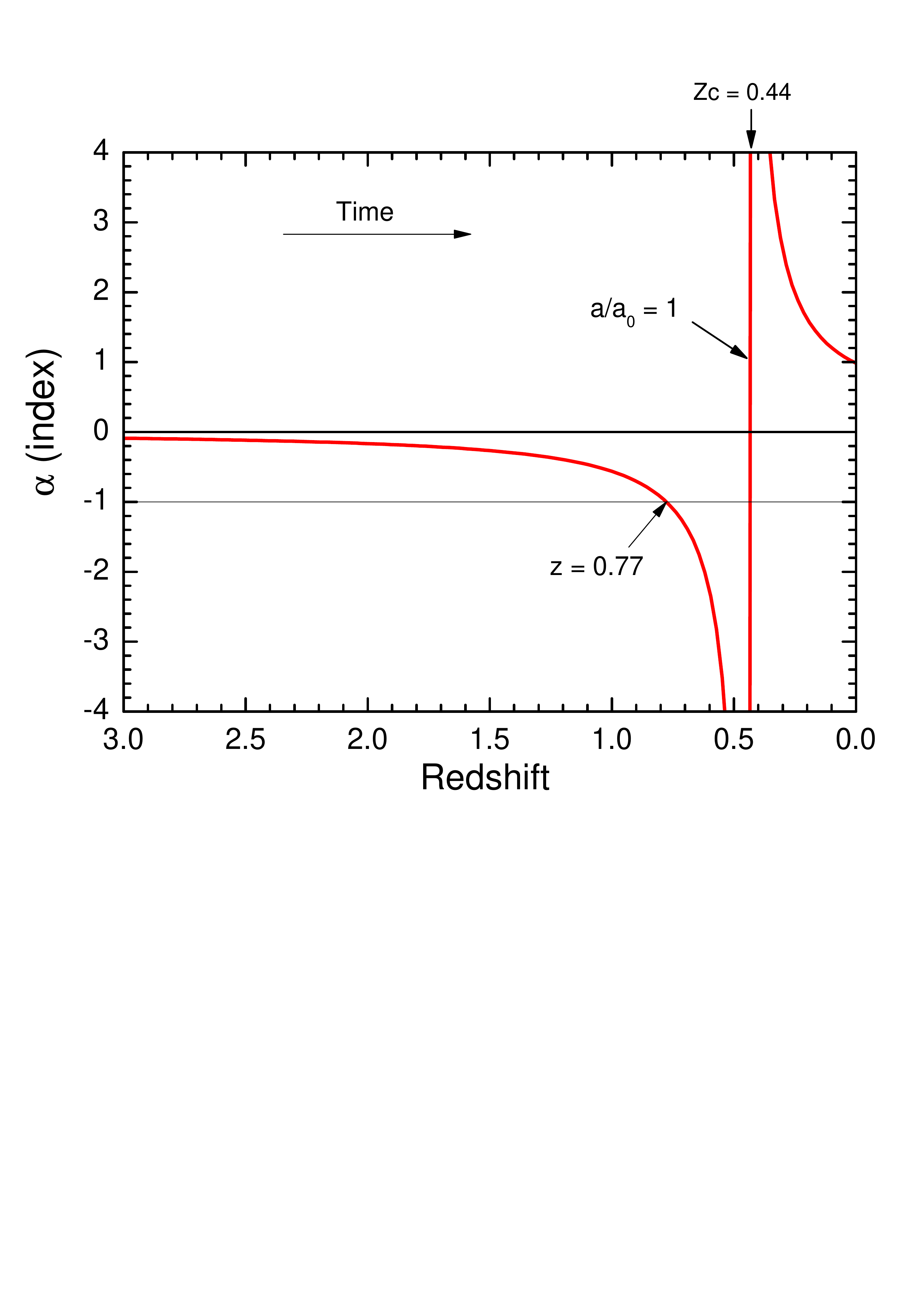}
	\vspace{-7.0cm}
    \caption{The index $\alpha$ as a function of the redshift. $\alpha$ is the index of the Deby first function when parametrizing as a power law as a function of the acceleration.}
    \label{fig:alpha_z}
\end{figure}

These two assumptions allow obtaining the dependence of the index $\alpha$ with the redshift. The result is shown in Fig.~\ref{fig:alpha_z}. 

Equation~\ref{eq:egt2} allow to obtain the asymptotic rotation speed as function of the galaxy radius and taking into account the relation for the centripetal acceleration as $a=\mathrm{v}^2/R$, the rotation speed is
\begin{equation}
\mathrm{v}=(GM a_0^{\alpha})^{1/(2\alpha +2)}R^{\beta}, \;\text{with}\;
\beta=(\alpha-1)/2(\alpha+1)
\label{eq:speed}
\end{equation}

that has two asymptotic limits


$$\mathrm{v}=\left\{\begin{array}{rc}
\sqrt{\frac{GM}{R} },&\mbox{if}\quad \alpha =0 \;\;(T/T_D > 1),\\
(GMa_0)^{1/4} , &\mbox{if}\quad \alpha=1 \;\;(T/T_D<1).
\end{array}\right.
$$

The first asymptotic limit happens at high temperatures, and DGT coincides with the deep-Newtonian regime, and this means Keplerian rotation curves,
and according to figure~\ref{fig:alpha_z},  the $\alpha=0$ index corresponds at high redshifts, that is, for distant galaxies, $z>4$.

The second asymptotic limit corresponds at low temperatures, and DGT  coincides with the Tully-Fisher relation. That is the deep-MOND regime, and we have the asymptotically flat rotation curves, and according to 
figure~\ref{fig:alpha_z} the $\alpha=1$ index correspond at very low redshifts,  for nearby galaxies, those with redshift close to zero.  
\begin{figure}
    \vspace{-0.0cm}
    \centering
	\includegraphics[width=12cm]{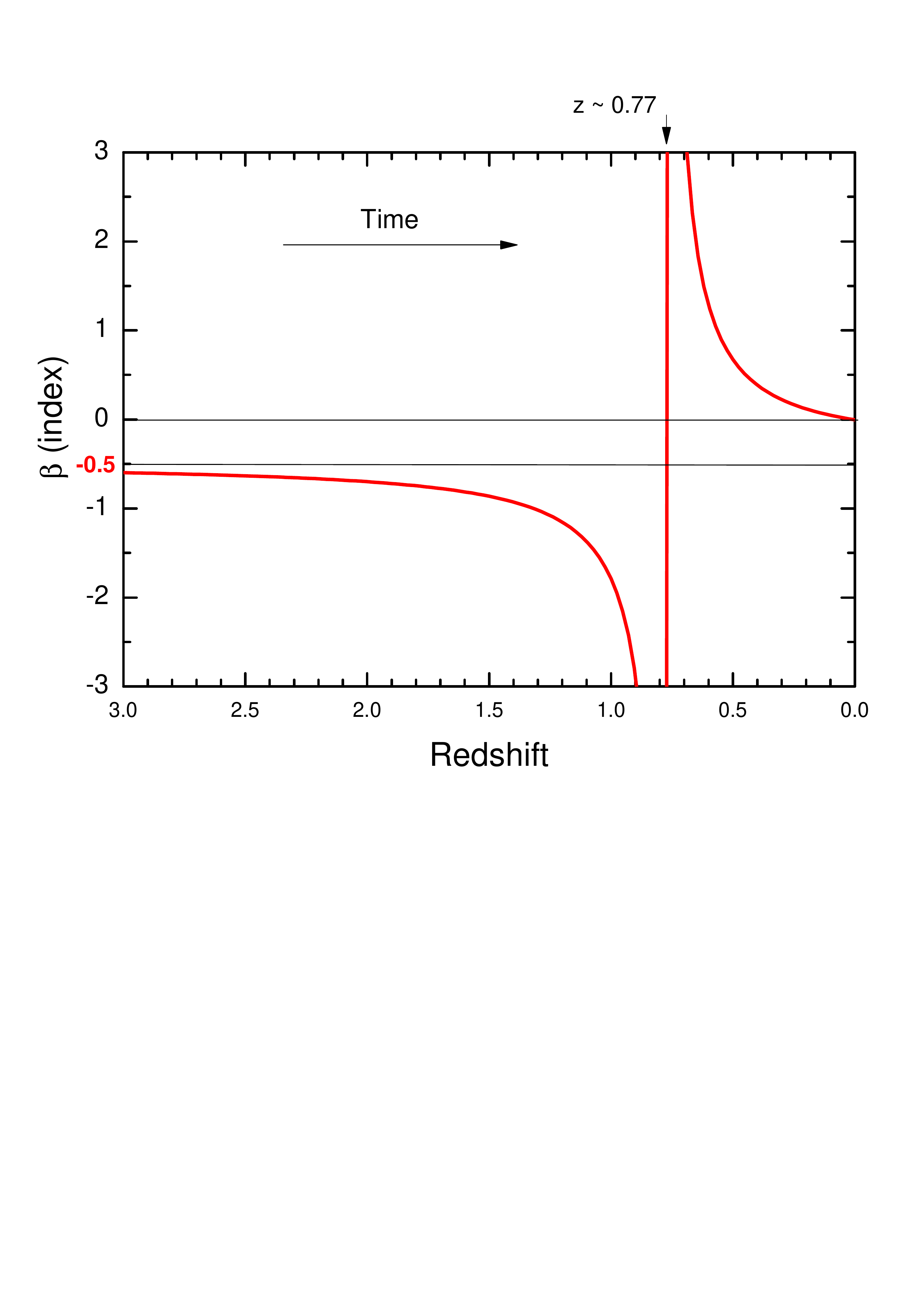}
	\vspace{-7.0cm}
    \caption{The index $\beta$ as a function of the redshift. $\beta$ is the exponent of the galaxy's radius in the expression of the rotation speed according to the DGT.
}
\label{fig:beta_z}
\end{figure}

The figure~\ref{fig:alpha_z} is also useful to help to understand the behavior of the rotation speed outside these two asymptotic limits.
Let's begin the analysis in $z=0.44$, at this redshift the index $\alpha$ tends to $\pm \infty$. But according to equation~\ref{eq:speed} the index $\beta$ tends to 0.5 to $\alpha=\pm \infty$, this means that the velocity of rotation is a continue function at $\alpha=\pm \infty$.
However, at $z\sim 0.77$ there is a discontinuity to $\alpha=-1$ and according to equation~\ref{eq:speed} the index $\beta$ tends to $\rightarrow \pm \infty$. 
Thus the rotation speed is not a continuous function of the redshift, diverge at  $z\sim 0.77$. This behavior can be seen better in figure~\ref{fig:beta_z} and represent the dependence of the index $\beta$ with the redshift.

From figure~\ref{fig:beta_z} we can see that that the transition between the two asymptotic limits is divided in two well defined regions around $z\sim 0.77$. The first from $z\leq 4$ to $z>0.77$, in this region the index $\beta$ is negative ranging from $\beta=-0.5$ (Keplerian rotation) to $\beta \rightarrow -\infty$ (stops rotation). 
The second region is from $z<0.77$ to $z=0$ in this region the index $\beta$ is positive ranging from $\beta \rightarrow \infty$ to $\beta=0$ (flat rotation curve).

\begin{figure}
    \vspace{-0.0cm}
    \centering
	\includegraphics[width=16cm]{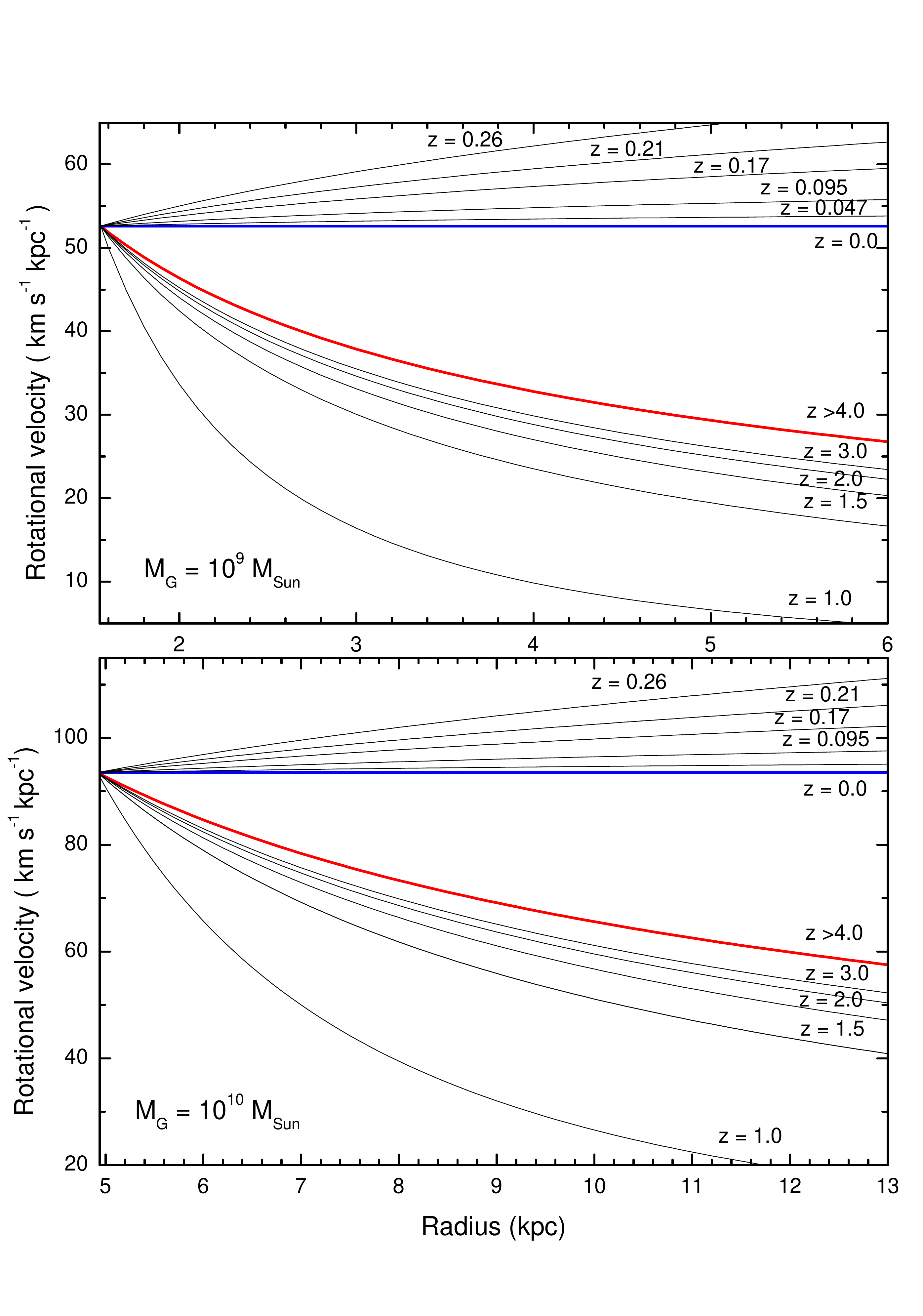}
	\vspace{-1.0cm}
    \caption{Expected galactic asymptotic rotation curves according to DGT, assuming two galaxies mass  $M_G=10^9 \times M_{Sun}$ top panel and  $M_G10^{10} \times M_{Sun}$  bottom panel. In both cases, there are two groups of galaxies divided according to their redshifts.  (a) The bottom group, those with redshift above $0.77$, with declining rotation curves like $R^{\beta}$ with $\beta\leqslant -0.5$ (Newtonian regime), in this case, the falling of the rotation curves is faster as the redshift decreases , and at $z=0.77$ the rotation velocity tends to zero. The red bold curves indicate the deep-Newtonian regime like $R^{-0.5}$.
(b) The top group, those with redshift below 0.77, with rising rotation curves like $R^{\beta}$ with $\beta \geqslant 0$ (Mondian regime), in this case, the rise of the rotation curves is faster as the redshift increases , and at $z=0.77$ the rotation velocity tends to infinite. The blue bold curves indicate the deep-MOND regime like $R^0$ (flat rotation curve).
}
\label{fig:rotation_two}
\end{figure}

The expected asymptotic rotation curves according to DGT is presented in figure~\ref{fig:rotation_two}.
The top panel assumes a galaxy mass of $10^9 \times M_{Sun}$  and the bottom panel is a galaxy mass of $10^{10} \times M_{Sun}$. In both cases, the galaxies are  in two groups, according to their redshift: 

(a) Bottom group, those with redshift above 0.77 with declining rotation curves like 
$R^{\beta}$ wit $\beta \leqslant -0.5$ (Newtonian regime), and the asymptotic velocity of rotation decreases as the radius of the galaxy increase. This sector includes the asymptotic limit ($\alpha= 0$) and correspond to $\beta =-0.5$, i.e., the deep-Newtonian regime and is represented by the bold red curves. The falls of the rotation curves are faster as the redshift decrease, and the rotation speed tends to zero at $z=0.77$.

(b) Top group, those with redshift below 0.77 with rissing rotation curves
like $R^{\beta}$ with $\beta \geqslant 0$ (Mondian regime), and the asymptotic velocity of rotation increases as the radius of the galaxy increase. This sector includes the asymptotic limit ($\alpha= 1$) and correspond to $\beta =0$, i.e., the deep-MOND regime and is represented by the bold blue curves. The rise of the rotation curves are faster as the redshift increase, and the rotation speed tends to infinite at $z=0.77$.

Recently has been reported declining rotation curves of six galaxies at moderate high redshift \citep{gen17}, ranging from $z>0.9$ to $z<2.4$. The data comes from the ESO - European Southern Observatory (Cerro Paranal Chile). Also, there is in the ESO data a second sample of measurements of 97 galaxies whose redshifts are in the range from $z> 0.6$ to  $z < 2.7$ with an average value of $z=1.52$ with declining rotation curves \citep{lan17}. However, in this case, was not possible to obtain high-quality rotation curves of individual galaxies and only was obtained a mean normalized curve for the sample as a whole.  These declining rotation curves observations at moderate high redshift, contrast with the measurements of nearby galaxies, where the rotation curves flatten out.

The DGT predict that galaxies with redshift above 0.77 are in the Newtonian regime.
The ESO observation corroborates this DGT prediction. The DGT, in addition to predicting a decline in the rotation curves of the galaxies with redshift above 0.7 also predicts that the fall of the rotational curves is faster than $ R^{-1/2}$, i.e., it is more rapid as the redshift decreases, and the rotation goes to zero to redshift 0.77.  Thus only galaxies with a very high redshift will be described by a deep-Newton regime (Keplerian-like rotation curves, $ \propto R^{-1/2}$).

However, a question arises, what prompted the galaxies to rotate again when they hit redshift $\sim 0.77$? A possible answer to explain the jump in the rotation 
of the galaxies is indicated in the next section.

\section{Cosmological implications}
\label{sec:cosmo}

According to DGT, the shape of the rotation of the galaxies defines two physical regimes. The Newtonian at high temperatures, and the Mondian at low temperatures, with two asymptotic boundaries. At very high temperatures (high redshifts), corresponds the deep-Newton regime (Keplerian-like rotation curves $\propto R^{-1/2}$) and at very low temperatures (redshift close to zero), corresponds the deep-MOND regime (flat rotation curves). The transition between these two regimes happens at $\sim 0.77$ when a mysterious  Dirac-like twisting force boost again the rotation of the galaxies, inducing the transition between these regimes. 

On the other hand, the observations of supernovae of type 1A seem to indicate an expanding Universe at an increasing rate. This accelerated expansion could be driving by an unknown entity, called as dark energy and described by the cosmological constant \citep{per99,rie98}.  Thus, the expansion of the universe has been accelerating since it entered in the so-called dominant dark-energy era, at redshift $z\sim 0.58$ \citep{bar08}. 

We show that the mysterious twisting force that boosted again the rotation of the galaxies at $z\sim 0.77$, inducing the transition between regimes, could be correlated, or to be the same entity that accelerated the galaxies expansion such as indicated by the observation of the type 1A supernovae.

\begin{figure}
    \vspace{-1.0cm}
    \centering
	\includegraphics[width=12cm]{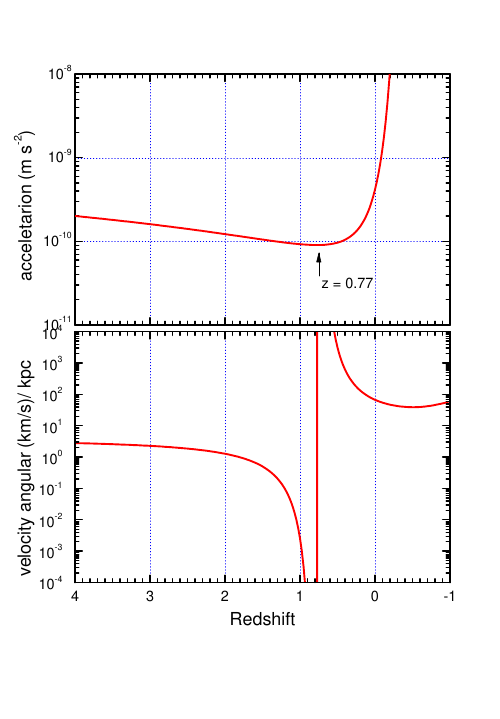}
	\vspace{-2cm}
    \caption{Top panel: the DGT prediction of accelerated expansion of galaxies, as a function of redshift. Bottom panel: the DGT prediction of the angular velocity for a mass galaxy $M=10^{10}\times M_{Sun}$, as a function of the redshift.
}
\label{fig:expantion}
\end{figure}

The inclusion of the thermal history of the universe in the DGT also allows obtaining the rate of change of the speed during the expansion, i.e., the acceleration as a function of the redshift. Introducing the Bekenstain entropy variation
$ \Delta S = 2 \pi k_B (mc/\hbar) \Delta X $ into the relation of the entropic force $ F = T \Delta S / \Delta X $, it is obtaining the relation
\begin{equation}
F=2\pi \frac{mc}{\hbar} k_B T.
\end{equation}
Following the analogy with the Debye theory, is obtains an expression to low temperatures substituting $k_BT$, for $k_BT\;\mathcal{D}_1(T_D/T)$
\begin{equation}
F=2\pi \frac{mc}{\hbar} k_B T_D (\frac{T}{T_D}) \mathcal{D}_1(T_D/T).
\end{equation}
Taking into account the results shown in section~\ref{sec:gravity} the Debye first function
can be written as
\begin{equation}
\mathcal{D}_1(T_D/T)=\mathcal{D}_1(a_0/a)=(a/a_0)^{\alpha},
\end{equation}
with the constrain condition $(a/a_0)=(\pi^2/6) \; (T/T_D)$ and $F=ma$,
the acceleration is estimated  to be
\begin{equation}
a=\frac{2\pi c k_B}{\hbar} T_D (\frac{T}{T_D}) \frac{6}{\pi^2}(a/a_0)^{\alpha}.
\end{equation}
The Debye temperature $T_D$ can be expressed as  function of the Milgrom acceleration
$a_0$ given by equation~\ref{eq:td} and considering the thermal history,
given by equation~\ref{eq:ratio} as 
$T/T_D=0.43(1+z)$, the acceleration is then
\begin{equation}
a=a_0\; [0.43(1+z)]^{1/(1-\alpha)}.
\end{equation}
Note that the index $\alpha$ defined in equation~\ref{eq:alpha} depend on also of the redshift, such as is shown in figure~\ref{fig:alpha_z}. Figure~\ref{fig:expantion} top panel, shows the DGT prediction to the acceleration of the galaxies expansion as a function of redshift. We can see that to $z>0.77$ the expansion is decelerating, however, starting from $ z\sim 0.77$, the rate of the velocity variation increases, that is, the expansion is accelerating. This transition from decelerating to accelerating expansion coincides with the transition to the rotation of the galaxies as shown in 
figure~\ref{fig:expantion} bottom panel. This coincidence means that the same entity
that accelerates the expansion also boosted the
rotation of the galaxies at $z\sim 0.77$.

The above results indicate that the expansion of the Universe is in continuous acceleration, starting at $z\sim 0,77 $, this contrasts with the acceleration of the rotation of the galaxies and that it is compatible with only a
 Dirac delta-like twisting force, around $z\sim 0.77$.

So far, the observation of type 1A supernovae is the strongest evidence on dark energy hypothesis. Because nowadays, in the Plank era, the CMB anisotropies spectrum provide limited information on the dark energy including the baryon acoustic oscillations (BAO). The same happens with other effects attributed to dark energy, such as the redshift-space distortions and the microlensing due to large scale structures.
Now we can see that the shapes of the galaxies rotation curves in the transition from high to low redshift ($z\sim 0.77$), it is also substantial evidence to the dark energy hypothesis. The DGT also provides the current ($z=0$) acceleration value
as $a \sim 4.1 \times 10^{-9} ms^{-2}T $. This value is in agreement with the observations.

A second signal for the transition to the rotation of the galaxies at $z\sim 0.77$ comes from their redshift distributions. The sudden jump in the rotation speed at $z\sim 0.77$ could reduce the size of the galaxies until they could no longer be detected,
 the effect would be a reduction in the number of galaxies around $z\sim 0.77$.  The large galaxy surveys corroborate this prediction, those that are extending for redshifts beyond $z>1$, that is, the so-called deep sky.

\begin{figure}
    \vspace{-1.0cm}
    \centering
	\includegraphics[width=12cm]{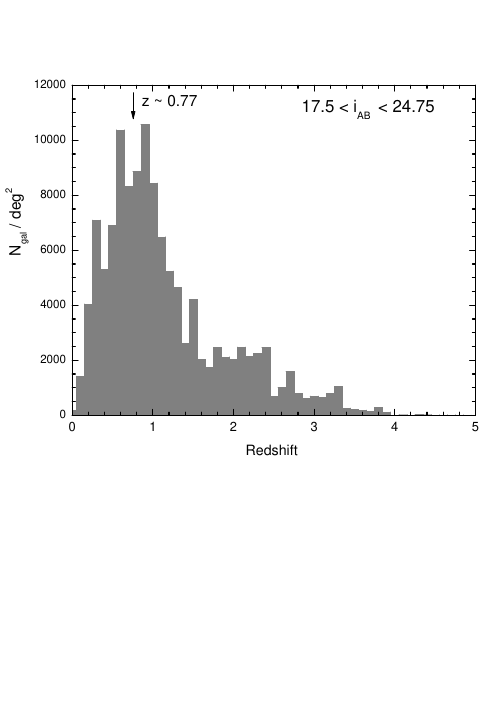}
	\vspace{-6cm}
    \caption{Redshift distribution of galaxies in the VIMOS VLT Deep Survey
     (L`{e}fevre et al., (2013)).
}
\label{fig:vlt}
\end{figure}

  The uncertainties on the redshifts of galaxies are at least 20\%; there are sources of uncertainty difficult to quantify, such as the gas and dust in the intergalactic space \citep{gwy96}.  Also, each survey has its experimental bias, different field of view and different selection criteria. 
  In general, the existence of peaks in the redshift distribution means that galaxies tend to cluster, forming large structures and the redshift can be affected by the contamination from star-forming galaxies.
Even under these limitations, some galaxy surveys show a signal of the rotation velocity transition. 

The first comes from the VIMOS VLT Deep Survey \citep{feb13} at ESO - European Southern Observatory (Cerro Paranal Chile) as is shown in figure~\ref{fig:vlt}. The VIMOS VLT Deep Survey population extends from $z = 0$ to $z\sim 4.0$ but is much more concentrated from $z=0.2$ to $z\sim 1.6$, peaking at around 0.75. However, the peak is no single, has a structure bimodal,  because there are two peaks and the central part of the gap between these peaks corresponds to  $z\sim 0.77$.

\begin{figure}
    \vspace{-0.0cm}
    \centering
	\includegraphics[width=12cm]{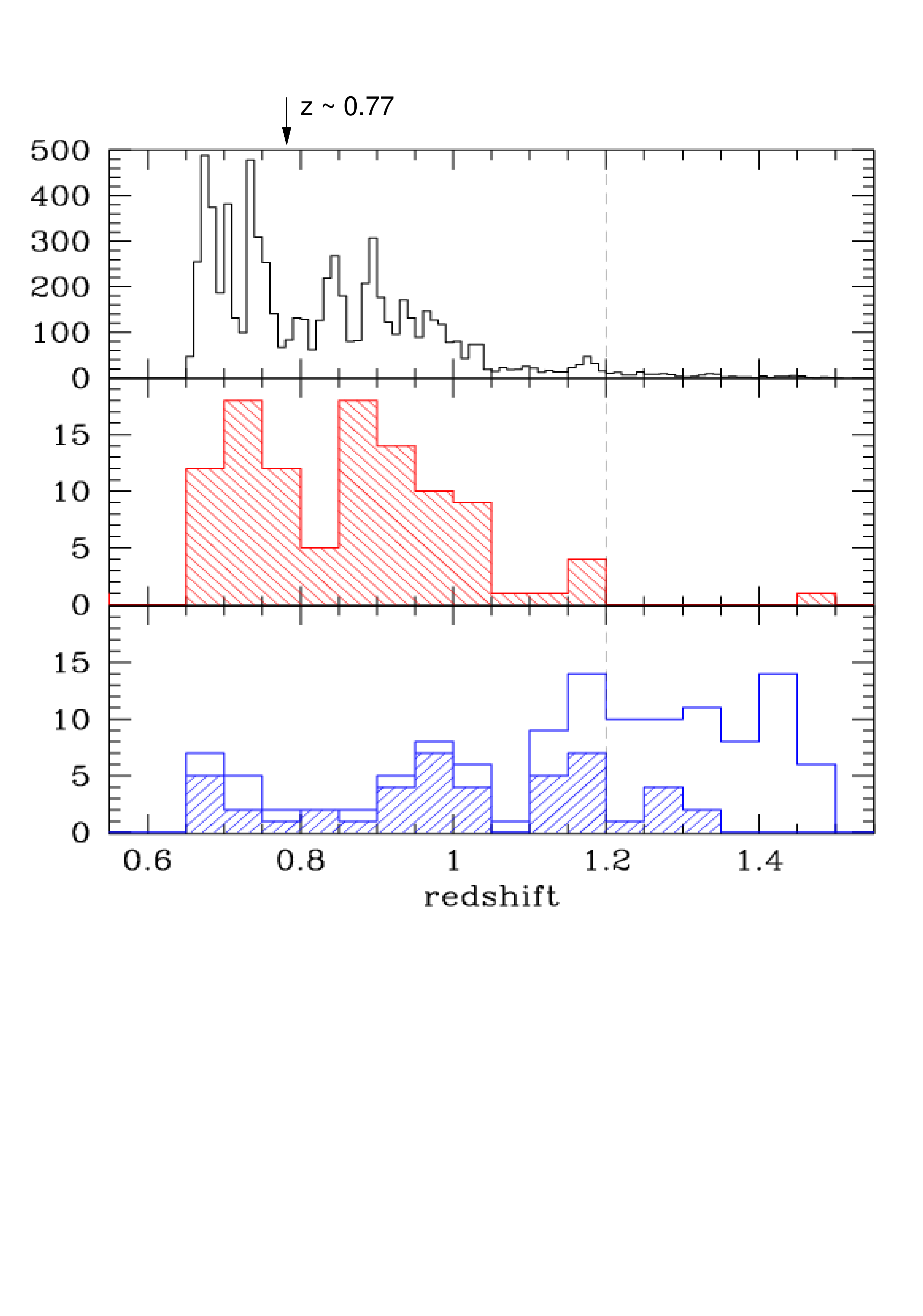}
	\vspace{-4cm}
    \caption{Top panel: Redshift Distributions of zCOSMOS-Bright Survey of extragalactic objects. Central panel:  narrow-emission-line galaxies with [Ne v] detection. Bottom panel:  broad-line AGN selected from the zCOSMOS survey (empty histogram); and the hatched histogram shows the [Ne v]-detected
objects. (Mignoli et al., (2013)).
}
\label{fig:zcosmos}
\end{figure}

There are also galaxies known as AGN (active galactic nuclei). Particularly,  those with emission line [Ne v] $\lambda 3426$. The [Ne v]-selected AGN sample, should not be affected by significant contamination from star-forming galaxies.
In the top panel of figure~\ref{fig:zcosmos} the redshift distributions of zCOSMOS-Bright Survey of extragalactic objects in the redshift range of ($0.65 < z < 1.50$) is plotted \citep{mig13}. The central panel represent the narrow emission-line galaxies with [Ne v] detection, and the bottom panel represent the broad-line AGN selected from the zCOSMOS survey (empty histogram), and the hatched histogram shows the [Ne v]-detected objects. We can see, that the signal of the transition of the rotation speed of the galaxies at $z\sim 077$ is evident in the [Ne v]-selected AGN sample, those less affected by the contamination from star-forming galaxies and plotted in the central panel.

\begin{figure}
    \vspace{-0.0cm}
    \centering
	\includegraphics[width=12cm]{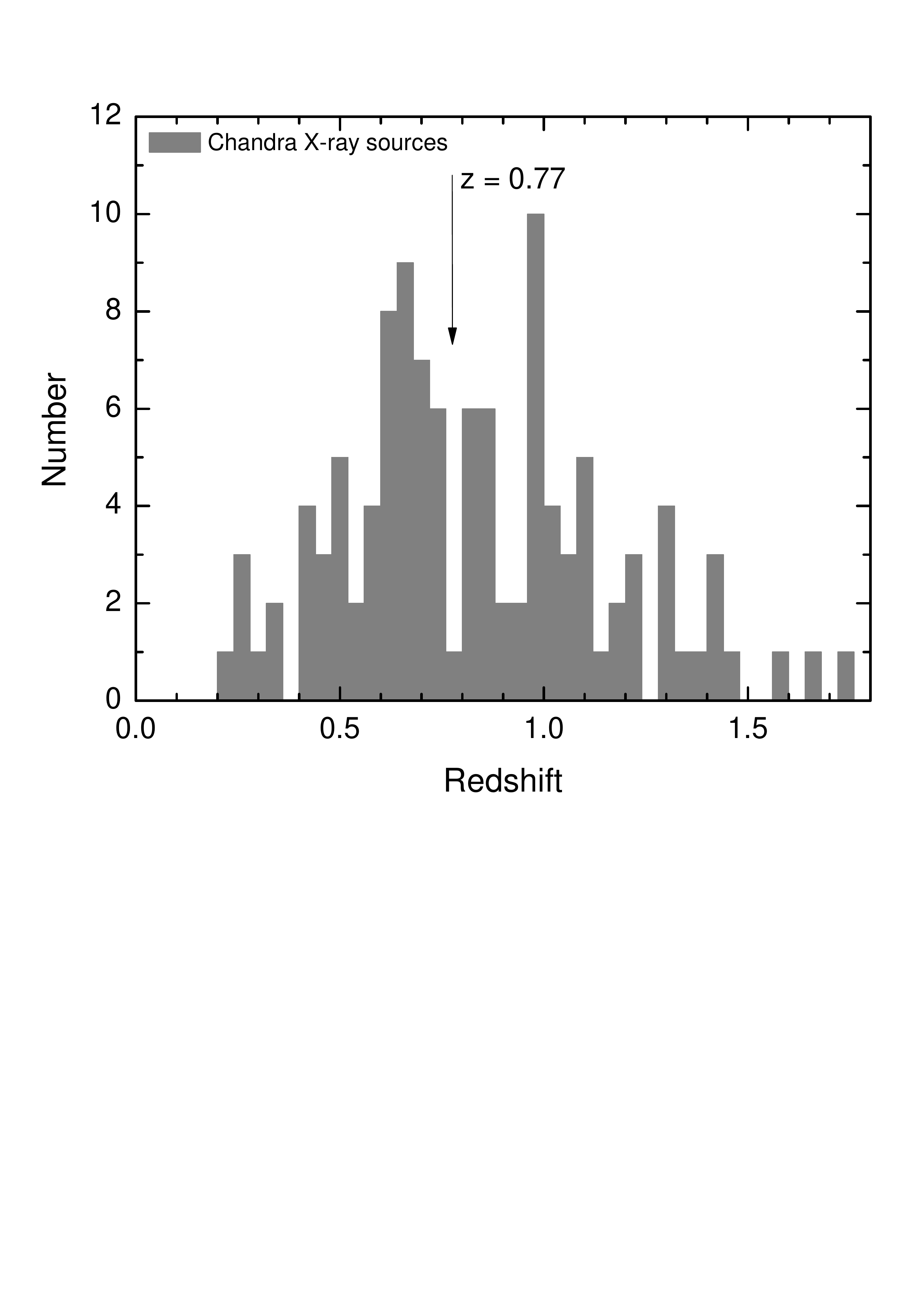}
	\vspace{-6cm}
    \caption{Redshift distribution of a sample of the AEGIS Chandra X-ray sources spectroscopically identified by cross-correlating with the DEEP2 photometric and redshift catalogs (Montero-Dorta et al.( 2009)).
}
\label{fig:chandra}
\end{figure}

We have also found a good signal of the transition in a sample of the AEGIS Chandra X-ray sources spectroscopically identified by cross-correlating with the DEEP2 photometric and redshift catalogs \citep{mon09}  and plotted in figure~\ref{fig:chandra}.

We emphasize that dark energy is not an input parameter in our analysis. The dark energy seems as responsible for the transition of the rotation velocity of the galaxies at $z\ 0.77 $, similarly as the dark energy seems as responsible for the accelerating expansion, observed in the analysis of supernovae 1A.

\section{Conclusions}
\label{sec:conclusion}

We have presented a straightforward analysis of the rotation dynamics of the galaxies based on the DGT. Starting with a brief description of the DGT main ingredients. The DGT provides its connection between acceleration and temperature, and it allows that the first Debye function which driven the transition from high to low temperatures be parametrized by a power law function. This thermodynamic description of gravity is useful because it allows to include the thermal history of the Universe, for example, in the relation provided by the DGT for the speed of rotation of the galaxies.

According to the DGT, the expression for the rotation of the galaxies is not a continuous function of the redshift. The discontinuity happens at $z\sim 0.77$ giving a sudden jump to very high rotation speeds and this behavior marks the transition from the Newtonian to Mondian regimes, at the expense of an extra boost. The redshift $z\sim 077$ also coincides with the start of the accelerated expansion. Indicating that a Dirac delta-like twisting force giving a boost to the rotation is the same entity which continuously accelerates the expansion. This last already observed using type 1A supernovae giving support to the dark energy hypothesis and described by the cosmological constant in the 
$ \Lambda CDM $ theory.

As already commented in section~\ref{sec:rotation} recently has been reported declining rotation curves of distant galaxies at moderate high redshift with an average redshift of $z=1.52$ \citep{lan17}  and in some individual galaxies in the range from $z\sim 0.9$ to $z\lesssim 2.4$ \citep{gen17}.  An interpretation according to VLT Group suggests that these galaxies have a mass fraction of baryons about the total dark matter halo, dominated by baryonic matter. 

According to the DGT, the observation of galaxies with redshift above 0.77 means observing galaxies before the transition from the Newtonian to the Mondian regimes induced by the dark energy, i.e., the rotation of galaxies with redshift above 0.77 would still be described by the Newtonian regime. Thus the DGT can make several predictions besides the decline of its rotation curves, for example: (a) the fall of the rotation curves of galaxies with redshifts above 0.77 is faster than $ R^{-1/2}$. (b) The rapidity of the fall of the rotation curves increases as the redshift decrease. (c) The deep-Newtonian regime described only very high redshift galaxies. The sample of galaxies with an average redshift of 1.52 \citep{lan17} in the VLT data, confirms the item (a).

While the observation of galaxies with redshifts less than 0.77 means that they have already crossed the transition induced by the dark energy from Newtonian to the Mondian regimes. Thus galaxies with redshift less than 0.77 are within the Mondian regime, and DGT can make several predictions, for example: 
(a) The fast rising of the rotation curves
decreases as the redshift decrease. (b) The observation of the deep-MOND regime is only in nearby galaxies, $z \sim 0$, (flat rotation curves).

Also, there is also a signal of the transition of the rotation speed of the galaxies at $ z \sim 0.77 $ in several galaxy surveys, including those cataloged as active galaxy nuclei (AGN). The signal is seen as a deficit in the number of galaxies around $ z \sim  0.77 $ and it is more evident in the samples that minimize the contamination from star-forming galaxies.

The DGT scenario for the rotation curve of galaxies together with the hypothesis of a flat universe allows obtaining the dark energy density. Taking into account that the energy density of the matter evolve as $(1+z)^3$ and considering that the critical energy density at $z=0$ is  $\rho_c=3.64\times 10^{-47}\; GeV^4$, with this normalization the energy density at $z=0.77$ correspond to the dark energy density and whose value is  $\rho_D = 1.78 \times 10^{-46} GeV^4$. Also, in DGT the current acceleration of the expansion ($z=0$) is $a=4.1\times 10^{-9} ms^{-2}$.

We highlight that the dark energy hypothesis is not assumed a priori but based on the results inferred from the analysis of the rotation velocity of the galaxies. The effects of the dark energy are within the theory; the dark energy is responsible for the transition between two physical regimes predicted by DGT.
Thus the DGT contrasts with other research on emergent gravity, such as the \citep{ver16} and until with the $\Lambda CDM$ theory, where the dark energy hypothesis is assumed a priori, i.e., they need of a cosmological constant already at the start of the analysis.

 Finally, we believe to be possible to improve the DGT, the discussion till now takes into account only the harmonic approximation. For instance, this limitation in the Debye theory of solids, lead at some inconsistencies whose consequences are: (a) there is no thermal expansion of solids and (b) the thermal conductivity of solids tends to infinite. One way to avoid these inconsistencies is to include higher order terms in the potential energy, that is, to include anharmonic terms in the phonon vibrations.
The inclusion of anharmonic terms in the DGT  maybe let us do a better description of astrophysical observations and even to make some new predictions, and this inclusion will be our next goal.

This work is supported by the National Council for Research (CNPq)
of Brazil, under Grants 308025/2012-1 and 307727/2015-7.


\end{document}